\documentclass[a4paper,10pt,twoside,twocolumn]{article}

\usepackage{header-private}

\newcommand\namecode[1]{{\textit{#1}}}

\begin{document}

\title{\vspace*{-0.75 cm} 
PDE-NetGen 1.0: from symbolic PDE representations of physical processes to trainable neural network representations.}
\author{O.~Pannekoucke$^1$~\footnote{olivier.pannekoucke@meteo.fr} and 
R.~Fablet$^2$\\
$^1$INPT-ENM, CNRM, CERFACS, 42, av. G. Coriolis 31057 Toulouse, France.\\
$^2$IMT-Atlantic, Lab-STICC, Brest, France.
}

\maketitle

\begin{abstract}
Bridging physics and deep learning is a topical challenge. While deep learning frameworks open avenues in physical science, the design of physically-consistent deep neural network architectures is an open issue. In the spirit of physics-informed NNs, \namecode{PDE-NetGen} package provides new means to automatically translate physical equations, given as PDEs, into neural network architectures. \namecode{PDE-NetGen} combines symbolic calculus and a neural network generator. The later exploits NN-based implementations of PDE solvers using Keras. 
With some knowledge of a problem, \namecode{PDE-NetGen} is a plug-and-play tool to generate physics-informed NN architectures. They provide computationally-efficient yet compact representations to address a variety of issues, including among others adjoint derivation, model calibration, forecasting, data assimilation as well as uncertainty quantification.
As an illustration, the workflow is first presented for the 2D diffusion equation, 
then applied to the data-driven and physics-informed identification of uncertainty dynamics for the Burgers equation.

\end{abstract}

\section{Introduction}
Machine learning and deep learning receive a fast growing interest in geo-science to address open issues, including for instance sub-grid parmeterization, 


A variety of learning architectures have shown their ability to 
encode the physics of a problem, especially deep learning schemes which typically involve millions 
of unknown parameters, while the theoretical reason of this success remains a key issue 
\citep{Mallat2016PTRS}. A recent research trend has involved the design of lighter neural network (NN) architectures, like ResNets with shared weights \citep{He2016inproc}, while keeping similar learning performance. 
Interestingly, a ResNet can be understood as an 
implementation of a numerical time scheme solving a ODE/PDE \citep{Ruthotto2019JMIV,Rousseau2019JMIV}. Applications to learning PDEs from data 
have also been introduced \eg PDE-Net \citep{Long2017A,Long2018A}.
These previous works emphasize the connection between the underlying physics and the NN architectures. 

Designing or learning a NN representation for a given physical process remains a difficult issue.
If the 
learning fails, it may be unclear to know how to improve the architecture of the neural network. Besides, it seems irrelevant to run computationally-expensive numerical experiments on large-scale dataset to learn well-represented processes. The advection in fluid dynamics may be a typical example of such processes, which do not require complex non-linear data-driven representations.
Overall, one would expect to accelerate and make more robust the learning process by combining, within the same 
NN architecture, the known physical equations with the unknown physics.

From the geoscience point of view, a key question is to bridge physical representations and neural network ones so that we can decompose both known and unknown equations according to the elementary computational units made available by state-of-the-art frameworks (e.g., keras, tensorflow). In other words, we aim to translate physical equations into the computational vocabulary available to neural networks. \namecode{PDE-NetGen} addresses this issue for PDE representations, for which we regard convolutional layers as being similar to the stencil approach, which results from a finite difference implementation of PDEs. \namecode{PDE-NetGen} relies on two main components: (i) a computer algebra system, here Sympy \citep{Meurer2017PCS}, used to handle the physical equations and discretize the associated spatial derivatives, (ii) a Keras network generator which 
automaticaly translate PDEs into neural network layers from these discretized forms. 
Note that code generator based on symbolic computation receives new interests to facilitate 
the design of numerical experiments see \eg \cite{Louboutin2019GMD}. As an illustration, we consider in this paper the application of \namecode{PDE-NetGen} to the identification of closure terms. 

The paper is organized as follows. In the next section, we detail the proposed neural network generator, 
with an illustration of the workflow on a diffusion equation. 
In section~\ref{sec3}, we present the numerical integration of the neural network implementation of the diffusion equation 
then an application to the data-driven identification of the closure of Burgers equation.  
Conclusion and perspective are given 
in section \ref{sec4}

\section{Neural Network Generatation 
from symbolic PDEs}\label{sec2}

Introducing physics in the design of neural network topology is challenging
since physical processes can rely on very different partial derivative
equations, \eg eigenvalue problems for waves or constrained
evolution equations in fluid dynamics under iso-volumetric assumption.  
The neural network code generator presented here focuses on physical processes given 
as evolution equations which writes 
\begin{equation}
 \pdt u = F(u,\partial^\alpha u),
\end{equation}
where $u$ denotes either a scalar field or multivariate fields, $\partial^\alpha u$ denotes 
partial derivatives with
respect to spatial coordinates, and $F$ is the 
generator of the dynamics. At first glance, this situation excludes diagnostic equation as 
encountered in geophysics, like balance equations: 
each equation has to be the evolution equation of a prognostic variable. \namecode{PDE-NetGen}
incorporates a way to solve diagnostic equation, this will be shown in the example detailed
in Section~\ref{sec3.2}.

We first explain how the derivatives are embedded into NN layers, then we detail the workflow 
of \namecode{PDE-NetGen} for a simple example.

\subsection{Introducing physical knowledge in the design of a NN topology}\label{sec2.1}

Since the NN 
generator is designed for evolution equations,
the core of the generator is the automatic translation
of partial derivatives with respect to spatial coordinates into layers. 
The correspondence between 
the finite-difference discretization and the convolutional layer 
give a practical way to translate a PDE into a NN 
\citep{Cai2012JAMS,Dong2017MMS,Long2017A}. 

For instance, the finite difference of a second order partial derivative 
$\pdx^2 u$, for $u(t,x)$ a one dimensional function, is given by 
\begin{equation}
\pdx^2 u(t,x) \approx 
	\frac{u(t,x+\delta x)+u(t,x-\delta x)-2u(t,x)}{\delta x^2},
\end{equation}
where $\delta x$ stands for the discretization space step.
This makes appear a kernel stencil 
$k=[1/\delta x^2,-2/\delta x^2,1/\delta x^2]$ that can be used in a 
1D convolution layer with a linear activation function and without bias.
A similar routine applies for 2D and 3D geometries. \namecode{PDE-NetGen} relies on 
the computer algebra system \namecode{sympy} \citep{Meurer2017PCS} to compute the stencil
as well as to handle symbolic expressions. Alternatives using automatic differentiation
can be considered as introduced by \cite{Raissi2018JMLR} who used TensorFlow for the 
computation of derivative.

Then, the time integration can be implemented either by a solver 
or by a ResNet architecture of a given time scheme \eg an Euler scheme or 
a fourth order Runge-Kutta (RK4) scheme \citep{Fablet2017A}.

These two components, namely the translation of partial derivatives into NN layers and a ResNet implementation of the time integration, are the building blocks of the proposed NN topology generator as examplfied in the next Section.


\subsection{Workflow of the NN representation generator}\label{sec2.2} 

\begin{figure}
\begin{center}
\includegraphics[width=8.3cm]{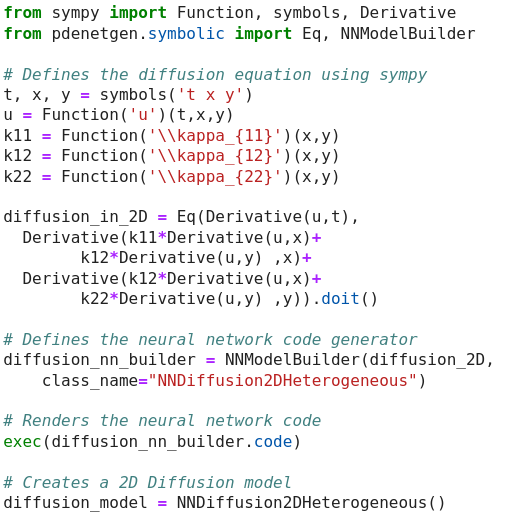}
\end{center}
\caption{\label{Fig:diffusion-code}
Neural Network 
generator for a heterogeneous 2D diffusion
equation}
\end{figure}

We now present the workflow for the NN 
generator 
given a symbolic PDE using the heterogeneous 2D diffusion equation as a testbed: 
\begin{equation}\label{Eq:diffusion}
\pdt u = \nabla\cdot\left(\bm{\kappa}\nabla u\right),
\end{equation}
where $\bm{\kappa}(x,y)$ is a field of $2\times 2$ tensors ($(x,y)$ are the spatial coordinates)
and whose python implementation is detailed in \Fig{Fig:diffusion-code}. 

Starting from a list of coupled evolution equations given as a PDE, 
a first preprocessing of the system determines the prognostic functions,
the constant functions, the exogenous functions and the constants. 
The exogenous functions are the functions which depends on time and space, 
but whose evolution is not described by the system of evolution equations. 
For instance, a forcing term in a dynamics is an exogenous function.

For the diffusion equation \Eq{Eq:diffusion},
the dynamics is represented in \namecode{sympy} using \namecode{Function}, 
\namecode{Symbol} and \namecode{Derivative} classes.
The dynamics is defined as an equation using the \namecode{Eq} class of \namecode{PDE-NetGen},
which inherits from the \namecode{Eq} class of \namecode{sympy}
with additional facilities (see the implementation in \Fig{Fig:diffusion-code} for additional details).

The core of the NN 
generator is given by the \namecode{NNModelBuilder} class. This class 
first preprocesses the system of evolution equations and translates the system into a python 
NN model.

The preprocessing of the diffusion equation \Eq{Eq:diffusion} 
presents a single prognostic function, $u$, and three 
constant functions $\kappa_{11},\kappa_{12}$ and $\kappa_{22}$. There is no
exogenous function for this example. During the preprocessing, the coordinate system 
of each function is diagnosed such that we may determine 
the dimension of the problem.
For the diffusion equation \Eq{Eq:diffusion}, since the function $u(t,x,y)$ is a function of $(x,y)$
the geometry is two-dimensional. 
In the current version of \namecode{PDE-NetGen}, only periodic boundaries are considered. 
The specific \namecode{DerivativeFactory} class ensures 
the periodic extension of the domain, then the computation of the derivative by using CNN 
and finally the crop of the extended domain to return to the initial domain.
Other boundaries could also be implemented and might be investigated in future developments.

\begin{figure*}[t]
\begin{center}
\includegraphics[width=12cm]{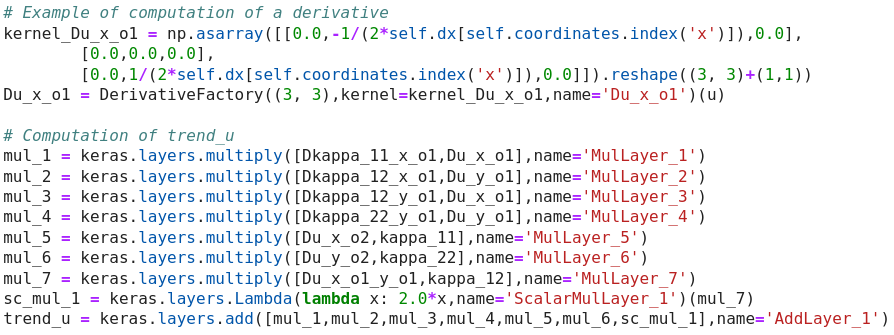}
\end{center}
\caption{\label{Fig:diffusion-keras} Part of the python code of the
 \namecode{NNDiffusion2DHeterogeneous} class  
which implements the diffusion equation \Eq{Eq:diffusion} as a neural-network by using Keras (only
one derivative is explicitly given, for the sake of simplicity)}
\end{figure*}

All partial derivatives with respect to spatial coordinates are detected
and then replaced by an intermediate variable in the system of evolution 
equations. The resulting system is assumed to be algebraic, which means that it only contains addition, 
subtraction, multiplication and exponentiation (with at most a real).
For each evolution equation, the abstract syntax tree is translated into a sequence of 
layers which can be automatically converted into NN layers in a given NN framework. For the current 
version of \namecode{PDE-NetGen}, we consider \namecode{Keras} \citep{Chollet2018book}. 
An example of the implementation in \namecode{Keras} is
shown in \Fig{Fig:diffusion-keras}: a first part of the code is used to compute all the derivatives using Conv layers of \namecode{Keras},
then \namecode{Keras} layers are used to implement the algebraic equation which represents the trend 
$\partial_t u$ of the diffusion equation \Eq{Eq:diffusion}.

\begin{figure}
\begin{center}
\includegraphics[width=8.3cm]{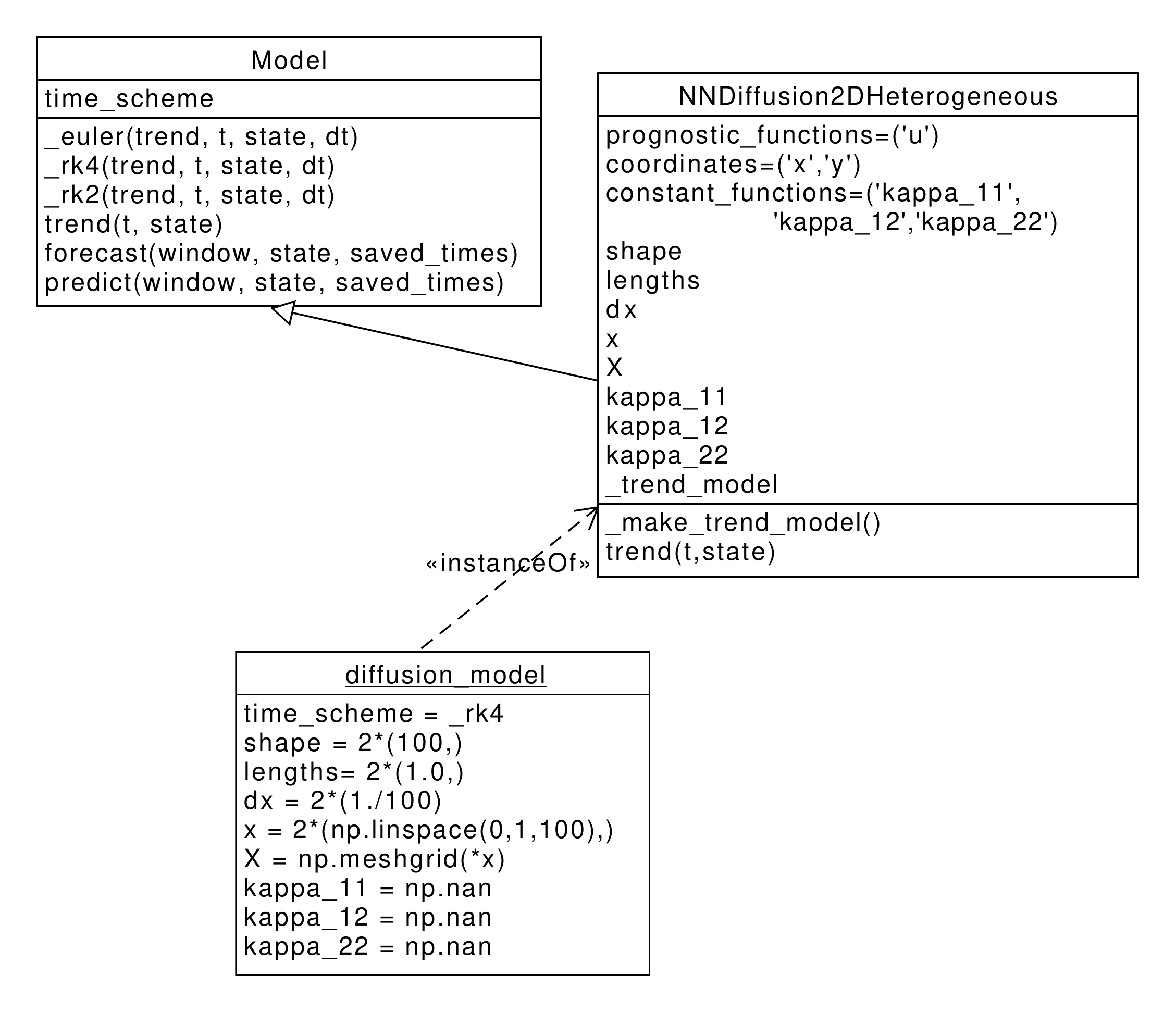}
\end{center}
\caption{\label{Fig:diffusion-uml}UML class diagram showing the interaction between the 
\namecode{Model} and the \namecode{NNDiffusion2DHeterogeneous} classes, 
and the resulting instance \namecode{diffusion\_model} 
corresponding to the numerical computation of the diffusion equation \Eq{Eq:diffusion}.}
\end{figure}

At the end, a python code is rendered from templates by using the \namecode{jinja2} package. 
The reason why templates are used is to facilitate the saving of the code in python modules 
and the modification of the code by the experimenter. Runtime computation of the class could be 
considered, but this is not implemented in the current version of \namecode{PDE-NetGen}.
For the diffusion equation \Eq{Eq:diffusion}, when run, the code rendered from the 
\namecode{NNModelBuilder} class 
creates the 
\namecode{NNDiffusion2DHeterognous} class. Following the class diagram \Fig{Fig:diffusion-uml}, 
the \namecode{NNDiffusion2DHeterogeneous} class inherits from a \namecode{Model} class which
implements the time evolution of an evolution dynamics by incorporating a time-scheme. Here several 
time-schemes are implemented, namely an explicit Euler scheme, a second and a fourth order Runge-Kutta scheme.

\section{Applications of \namecode{PDE-NetGen}}\label{sec3}

Two applications are now considered. First we validate the NN generator on a known
physical problem: the diffusion equation \Eq{Eq:diffusion} detailed in the previous section. Then, 
we tackle a situation where a part of the physics remains unknown, showing the benefit of merging 
the known physics in the learning of the unknown processes.

\subsection{
Application to the diffusion equation 
}\label{sec3.1}

In the python implementation \Fig{Fig:diffusion-code}, 
\namecode{diffusion\_model} is an instance of the \namecode{NNDiffusion2DHeterogeneous} class, 
which numerically solves the diffusion equation \Eq{Eq:diffusion} over a 2D domain, 
defined by default as the periodic domain $[0,1)\times[0,1)$ discretized by $100$ 
points along each directions, so that $dx=dy=1.0/100$.

\begin{figure*}[t]
\begin{center}
\includegraphics[width=12cm]{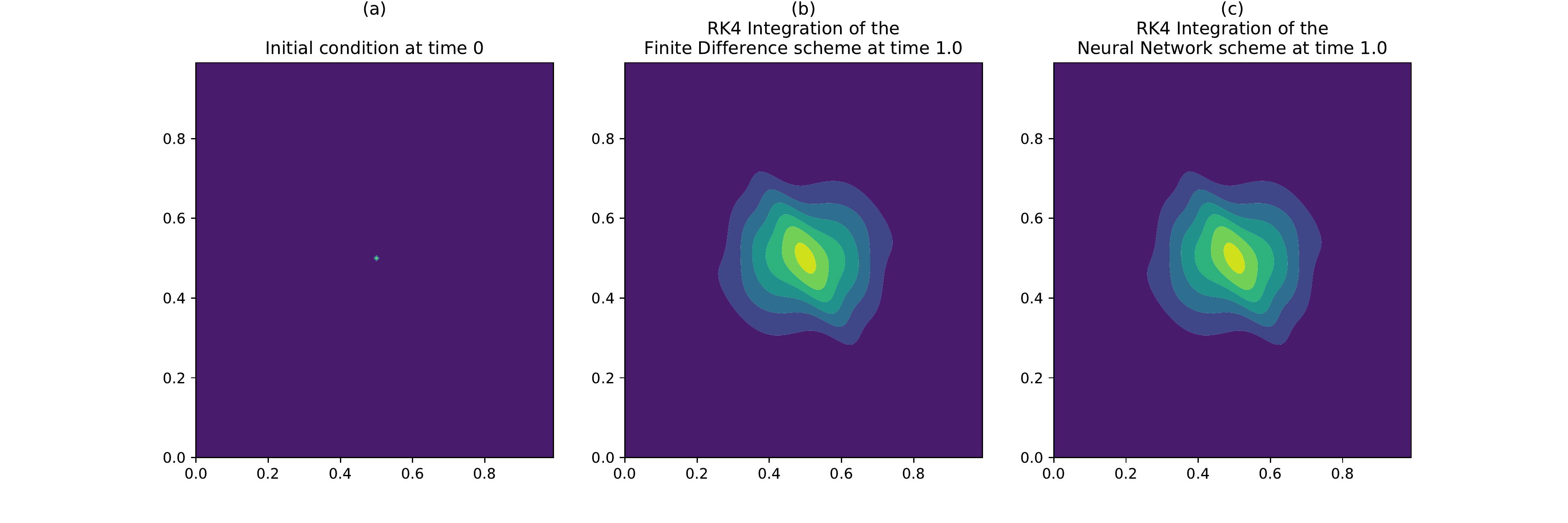}
\end{center}
\caption{\label{Fig:diffusion-num}
Starting from a Dirac (panel a), the diffusion equation \Eq{Eq:diffusion}
is integrated from $0$ to $1$ by using a fourth-order Runge-Kutta time scheme.
The results obtained from the time integration of the finite difference implementation 
\Eq{Eq:diffusion-fd} (panel b) and  of the generated NN representation (panel c) are similar.
}
\end{figure*}

The time integration of the diffusion equation is shown 
in \Fig{Fig:diffusion-num}. 
For this numerical experiment, the heterogeneous tensor field of diffusion
tensors $\bm{\kappa}(x,y)$ is set as rotations of the diagonal tensor $(l_x^2/\tau,l_y^2/\tau)$ defined from
the length-scales $l_x = 10\,dx$, $l_y=5\,dy$ and the time-scale $\tau = 1.0$, and with the 
rotation angles $\theta(x,y) = \frac{\pi}{3}\cos(k_x x+k_y y)$ where $(k_x,k_y)=2\pi(2,3)$. 
The time step for the 
simulation is $dt = \tau Min(dx^2/lx^2,dy/ly^2)/6$.
The numerical integration is computed by using a fourth-order Runge-Kutta scheme. 
The initial condition of the simulation is given by a Dirac \Fig{Fig:diffusion-num}~(a). 
In order to validate the solution obtained from the generated neural network, we compare 
the integration with the finite difference discretization of \Eq{Eq:diffusion}:
\begin{widetext}
\begin{small}
\begin{multline}\label{Eq:diffusion-fd}
\frac{\partial}{\partial t} u{\left(t,x,y \right)} 
= - \frac{2 \kappa_{22}{\left(x,y \right)} u{\left(t,x,y \right)}}{dy^{2}} + \frac{\kappa_{22}{\left(x,y \right)} u{\left(t,x,- dy + y \right)}}{dy^{2}} + \\
\frac{\kappa_{22}{\left(x,y \right)} u{\left(t,x,dy + y \right)}}{dy^{2}} + 
\frac{\kappa_{22}{\left(x,- dy + y \right)} u{\left(t,x,- dy + y \right)}}{4 dy^{2}} - \\
\frac{\kappa_{22}{\left(x,- dy + y \right)} u{\left(t,x,dy + y \right)}}{4 dy^{2}} - 
 \frac{\kappa_{22}{\left(x,dy + y \right)} u{\left(t,x,- dy + y \right)}}{4 dy^{2}} +\\
 \frac{\kappa_{22}{\left(x,dy + y \right)} u{\left(t,x,dy + y \right)}}{4 dy^{2}} + 
 \frac{\kappa_{12}{\left(x,y \right)} u{\left(t,- dx + x,- dy + y \right)}}{2 dx dy} - \\
 \frac{\kappa_{12}{\left(x,y \right)} u{\left(t,- dx + x,dy + y \right)}}{2 dx dy} -
 \frac{\kappa_{12}{\left(x,y \right)} u{\left(t,dx + x,- dy + y \right)}}{2 dx dy} + \\
\frac{\kappa_{12}{\left(x,y \right)} u{\left(t,dx + x,dy + y \right)}}{2 dx dy} + 
\frac{\kappa_{12}{\left(x,- dy + y \right)} u{\left(t,- dx + x,y \right)}}{4 dx dy} - \\
\frac{\kappa_{12}{\left(x,- dy + y \right)} u{\left(t,dx + x,y \right)}}{4 dx dy} -
 \frac{\kappa_{12}{\left(x,dy + y \right)} u{\left(t,- dx + x,y \right)}}{4 dx dy} +\\
 \frac{\kappa_{12}{\left(x,dy + y \right)} u{\left(t,dx + x,y \right)}}{4 dx dy} + 
\frac{\kappa_{12}{\left(- dx + x,y \right)} u{\left(t,x,- dy + y \right)}}{4 dx dy} -\\
 \frac{\kappa_{12}{\left(- dx + x,y \right)} u{\left(t,x,dy + y \right)}}{4 dx dy} - 
 \frac{\kappa_{12}{\left(dx + x,y \right)} u{\left(t,x,- dy + y \right)}}{4 dx dy} + \\
\frac{\kappa_{12}{\left(dx + x,y \right)} u{\left(t,x,dy + y \right)}}{4 dx dy} - 
\frac{2 \kappa_{11}{\left(x,y \right)} u{\left(t,x,y \right)}}{dx^{2}} +\\
 \frac{\kappa_{11}{\left(x,y \right)} u{\left(t,- dx + x,y \right)}}{dx^{2}} +
 \frac{\kappa_{11}{\left(x,y \right)} u{\left(t,dx + x,y \right)}}{dx^{2}} +\\
 \frac{\kappa_{11}{\left(- dx + x,y \right)} u{\left(t,- dx + x,y \right)}}{4 dx^{2}} -
 \frac{\kappa_{11}{\left(- dx + x,y \right)} u{\left(t,dx + x,y \right)}}{4 dx^{2}} - \\
\frac{\kappa_{11}{\left(dx + x,y \right)} u{\left(t,- dx + x,y \right)}}{4 dx^{2}} + 
\frac{\kappa_{11}{\left(dx + x,y \right)} u{\left(t,dx + x,y \right)}}{4 dx^{2}}
\end{multline}
\end{small}
\end{widetext}
which is shown in \Fig{Fig:diffusion-num}~(b). The heterogeneity of the diffusion tensors makes appear 
an anisotropic diffusion of the Dirac, which is perfectly reproduced by the result obtained
from the integration of the  generated neural network, shown in \Fig{Fig:diffusion-num}~(c). 
At a quantitative level, the $l^2$ distance between the both solutions is $10^{-5}$.
This validates the ability of the 
NN generator \namecode{PDE-NetGen} to compute the dynamics of a given physical evolution equation.

The next section illustrates the situation where only a part of the dynamics is
known, while the remaining physics is learned from the data. 

\subsection{Application to the data-driven identification of stochastic representations}\label{sec3.2}

As an illustration of the \namecode{PDE-NetGen} package, we consider a problem encountered
in uncertainty prediction: the parametric Kalman filter (PKF) 
\citep{Pannekoucke2016T,Pannekoucke2018NPG}. For a detailed presentation and discussion of uncertainty prediction issues in geophysical dynamics, 
we may refer the reader to \cite{Maitre2010book}. Here, we briefly introduce basic elements for the self-consistency of the example.


The idea of the PKF is to
mimic the dynamics of the covariance-error matrices all along the analysis and
the forecast cycle of the data assimilation in a Kalman setting
(Kalman filter equations for the uncertainty). It relies on  
the approximation of the true covariance matrices by some parametric covariance model. 
When considering a covariance model based on a diffusion equation, 
the parameters are the variance $V$ and the local diffusion tensor $\bm{\nu}$. Therefore, the
dynamics of the covariance-error matrices along the data assimilation cycles
is deduced from the dynamics of the variance and of the diffusion tensors. In place of the full covariance
evolution this dramatically reduces the dynamics to the one of few parameters.

For the non-linear advection-diffusion equation, known as the Burgers equation,
\begin{equation}\label{Eq:burgers}
\pdt u +u\pdx u = \kappa\pdx^2 u,
\end{equation}
the dynamics of the variance $\operatorname{{V_{u}}}$ 
and the diffusion tensor $\bm{\nu}_u=[\operatorname{{\nu_{u,xx}}}]$ 
(which is featured by a single field $\operatorname{{\nu_{u,xx}}}$),
writes \citep{Pannekoucke2018NPG}
\begin{equation}\label{Eq:NN-PKF}
\left\{
\begin{array}{lcl}
\frac{\partial}{\partial t} u &=& \kappa \frac{\partial^{2}}{\partial x^{2}} u - u \frac{\partial}{\partial x} u - \frac{  \frac{\partial}{\partial x} \operatorname{{V_{u}}} }{2}  \\
\frac{\partial}{\partial t} \operatorname{{V_{u}}} &=& - \frac{\kappa \operatorname{{V_{u}}}}{ \operatorname{{\nu_{u,xx}}} } + \kappa \frac{\partial^{2}}{\partial x^{2}} \operatorname{{V_{u}}} - \frac{\kappa \left(\frac{\partial}{\partial x} \operatorname{{V_{u}}}\right)^{2}}{2 \operatorname{{V_{u}}}} \\
	&&-  u  \frac{\partial}{\partial x} \operatorname{{V_{u}}} - 2 \operatorname{{V_{u}}} \frac{\partial}{\partial x} u  \\
\frac{\partial}{\partial t} \operatorname{{\nu_{u,xx}}} &=& 4 \kappa \operatorname{{\nu_{u,xx}}}^{2} \E{\operatorname{{\varepsilon_{u}}} \frac{\partial^{4}}{\partial x^{4}} \operatorname{{\varepsilon_{u}}}} \\
	&&- 3 \kappa \frac{\partial^{2}}{\partial x^{2}} \operatorname{{\nu_{u,xx}}}- \kappa + \frac{6 \kappa \left(\frac{\partial}{\partial x} \operatorname{{\nu_{u,xx}}}\right)^{2}}{\operatorname{{\nu_{u,xx}}}}\\
	&& - \frac{2 \kappa \operatorname{{\nu_{u,xx}}} \frac{\partial^{2}}{\partial x^{2}} \operatorname{{V_{u}}} }{  \operatorname{{V_{u}}}  } +
\frac{\kappa  \frac{\partial}{\partial x} \operatorname{{V_{u}}}  \frac{\partial}{\partial x} \operatorname{{\nu_{u,xx}}}}{ \operatorname{{V_{u}}}  } +\\
	&& \frac{2 \kappa \operatorname{{\nu_{u,xx}}}  \left(\frac{\partial}{\partial x} \operatorname{{V_{u}}}\right)^{2}  }{  \operatorname{{V_{u}}}^{2}  } -  u  \frac{\partial}{\partial x} \operatorname{{\nu_{u,xx}}} +\\
	&& 2 \operatorname{{\nu_{u,xx}}}  \frac{\partial}{\partial x} u 
\end{array}
\right .
\end{equation}
where $\E{\cdot}$ denotes the expectation operator. For the sake of simplicity, in this system of PDEs, 
$u$ denotes the expectation of the random field and not the random field itself as in (\Eq{Eq:burgers}).


In this system of 
PDEs, the term 
$\E{\operatorname{{\varepsilon_{u}}} 
\frac{\partial^{4}}{\partial x^{4}} \operatorname{{\varepsilon_{u}}}}$
can not be determined from the known quantities $u, \operatorname{{V_{u}}}$
and $\operatorname{{\nu_{u,xx}}}$. This makes appear a closure problem, {\em i.e.} determinining the unknown term as a function of the known 
quantities.
A naive assumption would be to consider a zero closure ($\text{closure}(t,x)=0)$. 
However, while the tangent-linear 
evolution of the perturbations along the Burgers dynamics is stable, 
the dynamics of the diffusion coefficient $\operatorname{{\nu_{u,xx}}}$ would lead to unstable dynamics as the coefficient of the second order term 
$-3\kappa\frac{\partial^{2}}{\partial x^{2}} \operatorname{{\nu_{u,xx}}}$ is negative. This  stresses further the importance of the unknown term to successfully predict the uncertainty.


Within a data-driven framework, one would typically explore a direct identification of the dynamics of diffusion coefficient $\operatorname{{\nu_{u,xx}}}$. Here, we exploit \namecode{PDE-NetGen} to fully exploit the known physics and focus on the data-driven identification of the unknown term 
$\E{\operatorname{{\varepsilon_{u}}}\frac{\partial^{4}}{\partial x^{4}} \operatorname{{\varepsilon_{u}}}}$ in the system of equations \Eq{Eq:NN-PKF}. It comes to replace term $\E{\operatorname{{\varepsilon_{u}}}\frac{\partial^{4}}{\partial x^{4}} \operatorname{{\varepsilon_{u}}}}$ in \Eq{Eq:NN-PKF} by an exogenous function 
\namecode{closure(t,x)} and then to follow the workflow detailed in Section~\ref{sec2.2}.



The unknown closure function is represented by a neural network 
(a Keras model) which implements the expansion
\begin{multline}\label{Eq:NN-closure}
\text{closure}(t,x) 
\sim 
a \frac{\frac{\partial^{2}}{\partial x^{2}} \operatorname{{\nu_{u,xx}}}{\left(t,x \right)}}{\operatorname{{\nu_{u,xx}}}^{2}{\left(t,x \right)}} + \\
b \frac{1}{ \operatorname{{\nu_{u,xx}}}^{2}{\left(t,x \right)}} +
c \frac{\left(\frac{\partial}{\partial x} \operatorname{{\nu_{u,xx}}}{\left(t,x \right)}\right)^{2}}{\operatorname{{\nu_{u,xx}}}^{3}{\left(t,x \right)}}  
\end{multline}
where $(a,b,c)$ are unknown and where the partial derivatives are computed from 
convolution layers, as described in Section~\ref{sec2}. 
This expression
is similar to a dictionary of possible terms as in \cite{Rudy2017SA}
and it is inspired from an arbitrary
theoretically-designed closure for this problem, 
where $(a,b,c)=(1,\frac{3}{4},-2)$ \citep{Pannekoucke2018NPG}.
In the NN implementation of the exogenous function modeled as 
\Eq{Eq:NN-closure}, 
each of the unknown coefficients $(a,b,c)$ are
implemented as a 1D convolutional layer, with a linear activation function and 
without bias.
Note that the estimated parameters $(a,b,c)$ could be different from the
one of the theoretical closure: while the theoretical closure 
can give some clues for the design of the unknown term, this closure is not 
the truth which is unknown.

For the numerical experiment, the Burgers equation is solved on 
a one-dimensional periodic domain of length $1$, discretized in $241$ points.
The time step is $dt=0.002$, and the dynamics is computed over $500$
time steps so to integrate from $t=0$ to $t=1.0$. 
The coefficient of the physical
diffusion is set to $\kappa=0.0025$.
The numerical setting considered for the learning is the tangent-linear regime described in 
\cite{Pannekoucke2018NPG} 
where the initial uncertainty is small and whose results are shown in their Fig.~4(a), Fig.~5(a) 
and Fig.~6(a). 

\begin{figure}[t]
\begin{center}
\includegraphics[width=8cm]{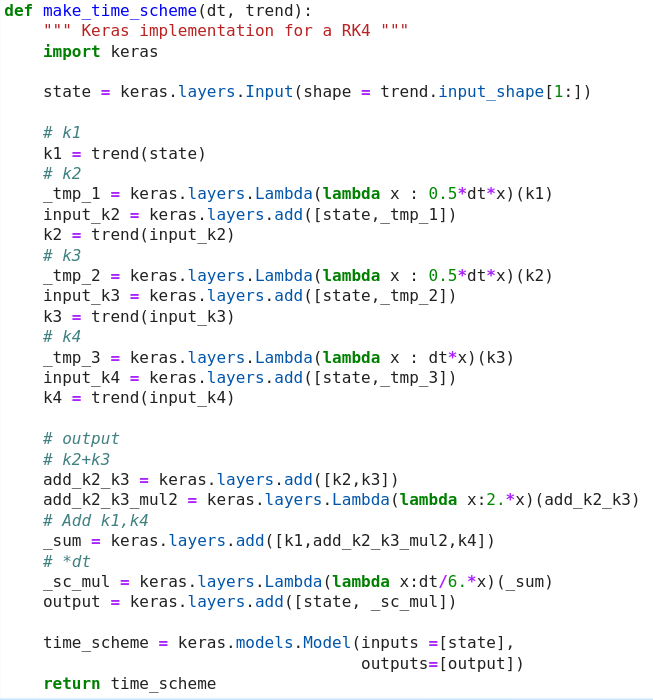}
\end{center}
\caption{\label{Fig:RK4}
Example of a Keras implementation for a RK4 time-scheme: given time-step $dt$ and a Keras model \namecode{trend} 
of the dynamics, the function \namecode{make\_time\_scheme} return a Kera model implementing a RK4.
}
\end{figure}

To train the parameters $(a,b,c)$ in \Eq{Eq:NN-closure}, 
we build a training dataset from an ensemble prediction method where each member is a numerical solution of the Burgers equation. The numerical code for the Burgers
equation derives from \namecode{PDE-NetGen} applied on the symbolic dynamics \Eq{Eq:burgers}. Using this numerical code, we generate 
a training dataset composed of $400$ ensemble simulations of $501$ time steps, where each each ensemble contains $400$ members. 
For each ensemble forecast, we estimate the mean,  
variance $\operatorname{{V_{u}}}$ 
and diffusion tensor $\bm{\nu}_u$. Here, we focus on the development of the front where we expect the unknown term to be of key importance and keep for training purposes the last $100$ time-steps of each ensemble forecast.
For the training only, the RK4 time-scheme is computed as the ResNet implementation given in Fig.~\ref{Fig:RK4}, 
so to provide the end-to-end NN implementation of the dynamics.

The resulting dataset 
involves $40000$ samples. To train the learnable parameters $(a,b,c)$, we minimize the one-step ahead prediction loss for the diffusion tensor $\bm{\nu}_u$. We use  ADAM optimizer \citep{Kingma2014A} and a batch size of $32$. Using an initial learning rate of $0.1$, the training converges within 3 epochs with a geometrical decay of the learning rate by a factor of $1/10$ after each epoch. 
The coefficients resulting from the training over $10$ runs are 
$(a,b,c)=( 0.93 ,  0.75, -1.80)\pm(5.1\,\, 10^{-5}, 3.6\,\, 10^{-4}, 2.7\,\, 10^{-4})$. 

\begin{figure*}[t]
\begin{center}
\includegraphics[width=12cm]{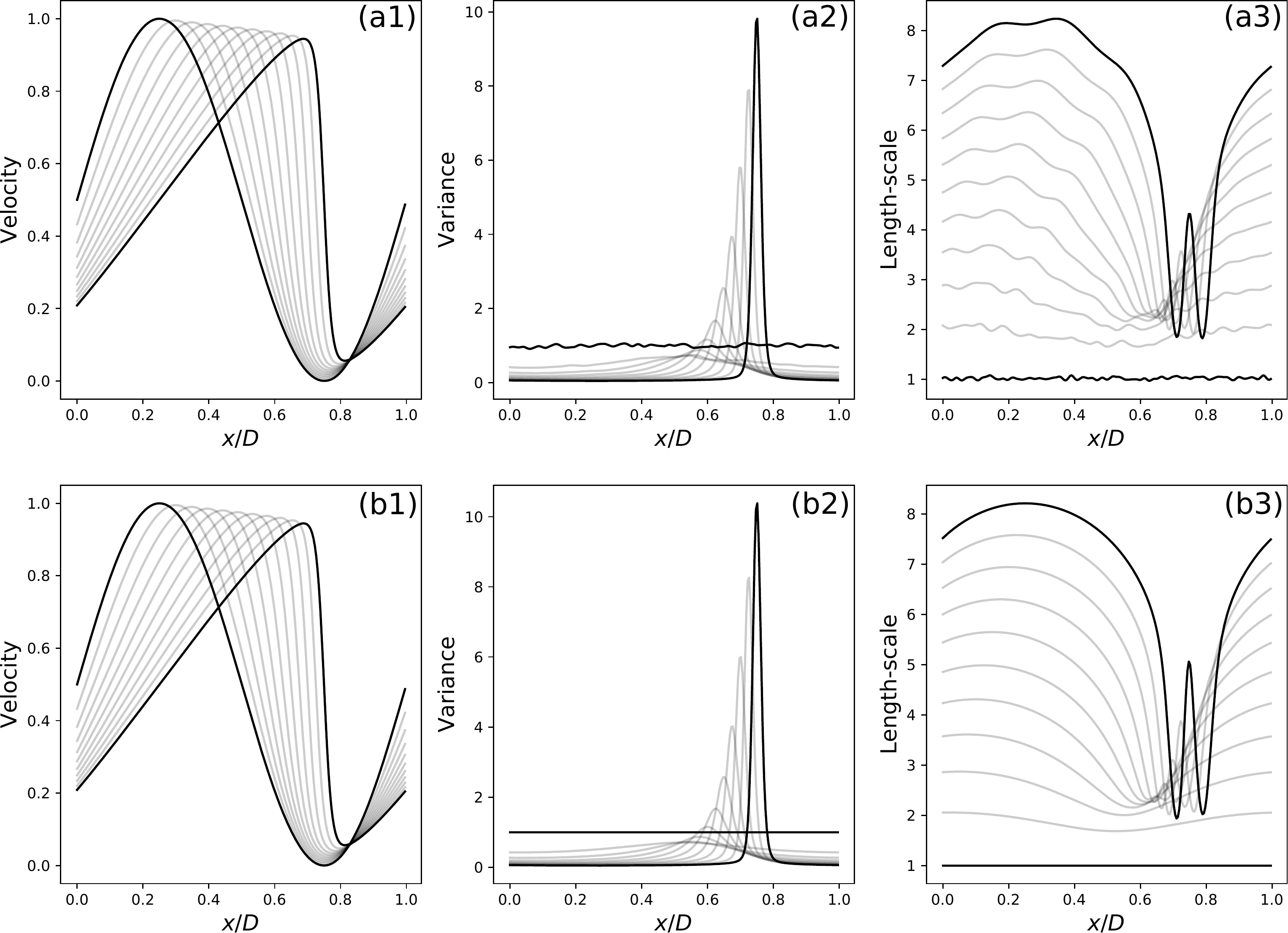}
\end{center}
\caption{\label{Fig:burgers}
Uncertainty estimated from a large ensemble of $1000$ members (a) with the 
expectation $\E{u}$ (a1), variance $\operatorname{{V_{u}}}$ (a2)
and the length-scale (defined from the diffusion coefficient by 
$\sqrt{0.5\operatorname{{\nu_{u,xx}}}}$) (a3) ; and the uncertainty predicted from the PKF evolution
equations closed from the data (b), where the same statistics are shown in (b1), (b2) and (b3).
The fields are represented only for time $t={0,0.2,0.4,0.6,0.8,1}$
}
\end{figure*}

Figure~\ref{Fig:burgers} compares the estimation from a large ensemble of $1000$ members (top panels)
with the results of the trained closed PKF dynamics (bottom panels).  Both the ensemble and PKF means (a1) and (b1) clearly show a front which emerges from the smooth initial 
condition and located near $x=0.75$ at time $1.$. 
The variance fields (a2) and (b2) illustrate the vanishing of the variance due to the 
physical diffusion (the $\kappa$ term in \Eq{Eq:burgers}) 
and the emergence of a pic of uncertainty which is related to the uncertainty of 
the front position. Instead of the diffusion  $\operatorname{{\nu_{u,xx}}}$, panels (a3) and (b3)
show the evolution of the correlation length-scale defined as 
$\sqrt{0.5\operatorname{{\nu_{u,xx}}}}$, which has the physical dimension of a length. Both panels show
the increase of the length-scale due to the physical diffusion, except in the vicinity of the front
where an oscillation occurs, which is related to the inflexion point of the front. 
While the magnitude of the oscillation predicted by the PKF (b3) is slightly larger 
than the estimation from the large ensemble reference (a3), 
the pattern is well predicted by the PKF. Besides, the parametric form of the PKF does not involve local variabilities due to the finite size of the ensemble, which may be observed in panel (a3).
Overall, these experiments support the relevance of 
the  closure \Eq{Eq:NN-closure} learned from the data to capture the uncertainty associated with Burgers' dynamics.

\subsection{Discussion on the choice of a closure}

In the Burgers' dynamics, an a priori knowledge was introduced 
to propose a NN implementing the closure \Eq{Eq:NN-closure}.

In the general case,  the choice of the terms to be introduced in the closure may be guided by known 
physical properties that need to be verified by the system. For example, 
conservation or symmetries properties that leave the system invariant 
can guide in proposing possible terms 
For the Burgers' dynamics, $\operatorname{{\nu_{u,xx}}}$ has the dimension 
of a length squared, $[L^2]$, and  
 $\E{\operatorname{{\varepsilon_{u}}}\frac{\partial^{4}}{\partial x^{4}} \operatorname{{\varepsilon_{u}}}}$ is of dimension
$[L^{-4}]$. Thus, the terms considered in \Eq{Eq:NN-closure} are among 
the simplest ones which fullfill the expected dimensionality of $[L^{-4}]$.
Symbolic computation may here help the design of such physical parameterizations in more general cases.

When no priors are available, one may consider modeling the 
closure using state-of-the-art deep neural network architectures 
which have shown impressive prediction performance, \eg CNNs, ResNets
\citep{Zagoruyko2016WRN,Raissi2018JMLR}.

\section{Conclusion}  
\label{sec4}

We have introduced a neural network generator \namecode{PDE-NetGen}, which provides new means to bridge physical priors given as symbolic PDEs and learning-based NN frameworks.  
This package derives and implements a finite difference version of a system of evolution equations,
where the derivative operators are replaced by appropriate convolutional layers 
including the boundary conditions. The package has been developed in python using the symbolic mathematics library \namecode{sympy} and \namecode{keras}.


We have illustrated the usefulness of \namecode{PDE-NetGen}
through two applications: a neural-network implementation of a 2{\sc d} heterogeneous diffusion equation and the uncertainty prediction in the Burgers equation. The later involves unknown closure terms, which are learned from data using the proposed neural-network framework.   
Both illustrations show the potential of such an approach, which could be useful for improving the 
training in complex application by taking into account the physics of the problem.

This work opens new avenues to make the most of existing physical knowledge and of recent advances in data-driven settings, and more particularly neural networks, for geophysical applications. This includes a wide range of applications, where such physically-consistent neural network frameworks could either lead to the reduction of the computational cost (e.g., GPU implementation embedded in deep learning frameworks) or provide new numerical tools to derive key operators (e.g., adjoint operator using automatic differentiation). Besides, these neural network representations also offer new means to complement known physics with the data-driven calibration of unknown terms. This is regarded as key to advance the state-of-the-art for the simulation, forecasting and reconstruction of geophysical dynamics through model-data-coupled frameworks.



\vskip 0.5cm
\textit{Code availability.} Code of PDE-NetGen is available on github in
version 1.0: https://github.com/opannekoucke/pdenetgen
Code of \namecode{PDE-NetGen} is available on github in version 1.0: 
\url{https://github.com/opannekoucke/pdenetgen}

\section*{Acknowledgments}
The UML class diagram has been generated from UMLlet \cite{Auer2003inproc}.
This work was supported by the French national
program LEFE/INSU (Étude du filtre de KAlman PAramétrique,
KAPA).RF has been partially supported by Labex Cominlabs (grant SEACS), 
CNES (grant OSTST-MANATEE) and ANR through programs EUR Isblue, 
Melody and OceaniX.
\includegraphics[scale=0.1]{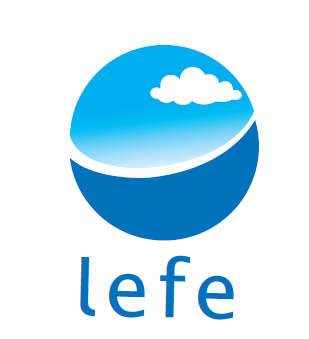}


\bibliographystyle{ametsoc}
\bibliography{arxiv-main}


\end{document}